\documentclass[12pt]{elsarticle}

\usepackage{graphicx}
\usepackage{amssymb}
\usepackage{amsmath, nccmath}
\usepackage{commath}
\usepackage{amsthm}
\usepackage{geometry}
\usepackage{lineno}

\usepackage{lipsum}
\makeatletter
\def\ps@pprintTitle{%
 \let\@oddhead\@empty
 \let\@evenhead\@empty
 \def\@oddfoot{}%
 \let\@evenfoot\@oddfoot}
\makeatother

\begin{document}

\begin{frontmatter}


\title{Convergence in divergent series related to perturbation methods using continued exponential and Shanks transformations}
 \author{Venkat Abhignan}
 \address{venkat.a@qditlabs.com, yvabhignan@gmail.com}
 \address{Qdit Labs Pvt. Ltd., Bengaluru - 560092, India}


\begin{abstract}

Divergent solutions are ubiquitous with perturbation methods. We use continued function such as continued exponential to converge divergent series in perturbation approaches for energy eigenvalues of Helium, Stark effect, and Zeeman effect on Hydrogen. We observe that convergence properties are obtained similar to those of the Pad\'e approximation, which is extensively used in literature. Free parameters are not used, which influence the convergence, and only the first few terms in the perturbation series are implemented.
\end{abstract}


\end{frontmatter}


\section{Introduction}
The vast majority of problems characterizing real-world systems cannot be exactly solved beyond the hydrogen atom in quantum physics. While laborious computations are occasionally possible, they are insufficient for comprehending the physics of the phenomenon under consideration. Perturbation theory in quantum physics, which involves studying a system's characteristic features through a series of approximations, is the most common method for addressing complex issues. However, the existence of small parameters suitable for perturbation theory is rare. Furthermore, the resulting perturbative series typically diverges even when the parameter is small. These challenges of lacking small parameters and divergent series necessitate the use of resummation techniques. 

While various techniques exist to create approximations of solutions \cite{bender1999advanced} and circumvent these divergent problems, the rigorous analysis by Stieltjes on continued fractions has led to the applicability of its analogue Pad\'e sequences on a wide range of eigenvalue problems of perturbation theory \cite{LOEFFEL1969656,chenpra,Vainberg1998,AUSTIN2,Silverstone_1979}. Recently, it was shown that even other continued functions like continued exponential have remarkably interesting convergence properties \cite{POLAND1998} by obtaining results related to phase transitions in quantum field theory \cite{abhignan2020continued, Abhignan_2021,Abhignan2023,Abhignan2023_2}. Poland compared the convergence advantages of continued exponentials with Padé approximants in a first-order phase transition model, noting that Padé methods can introduce spurious poles (singularities) that limit their effectiveness \cite{POLAND1998}. In certain problems related to second-order phase transition \cite{Abhignan2023_2}, continued functions exhibited superior convergence when directly compared with Padé approximants. To further illustrate these methods, we use continued exponential to get eigenvalues in standard textbook examples \cite{simon1991} of Helium (He) isoelectronic species, Stark effect, and Zeeman effect on Hydrogen (H) atom. The perturbation series for the energy levels is obtained in powers of a perturbation parameter. The inverse of the atomic number in the case of helium species, the magnetic field in the case of the Zeeman effect, and its analogue, the electric field in the case of the Stark effect. We convert these perturbation series into an asymptotic continued exponential \cite{contexp} to obtain a convergent eigenvalue. These series have zero radii of convergence in their perturbation series representation or are confined by the poles in other representations. Generally, such series can be recast into a continued function to obtain convergence outside this radius.

\section{Ground-state energy of He isoelectronic species}
Hylleraas used perturbation theory to solve the Hamiltonian of a two-electron atomic ion \cite{Hylleras} and obtained the nonrelativistic ground-state energy eigenvalues as the following perturbation series
\begin{equation}
    E(Z)=Z^2\sum_{i=0}^{\infty}\rho_i Z^{-i}
\end{equation}
(in atomic units (a.u.)) where Z is the nuclear charge. Accurate coefficients for this perturbation series were obtained by Hylleraas-Knight-Scherr variational perturbation method \cite{PhysRevA.41.1247}. 401 coefficients were used to compute the ground-state energy of the He atom as $-2.903 724 377 034 1167$ with precision of 30 decimal places. We convert this perturbation series into a continued exponential \begin{equation}
    Z^2\sum_{i=0}^{\infty}\rho_i Z^{-i} \sim Z^2a_0\exp(a_1Z^{-1} \exp(a_2 Z^{-1} \exp(a_3 Z^{-1} \exp(a_4 Z^{-1}\cdots))))
\end{equation}
with the same precision. The coefficients $a_i$ are obtained by expanding the continued exponential as a Taylor series, and relating them to the coefficients of the perturbation series up to order $i = 9$. 

Expanding any continued exponential \begin{equation} b_0\exp(b_1\lambda \exp(b_2 \lambda \exp(b_3 \lambda \exp(b_4 \lambda\exp(b_5 \lambda\cdots))))) \; \; \end{equation} as Taylor series and relating at each order of perturbation parameter $\lambda$ with a general power series \begin{equation}
    \sum_{i=0} \varepsilon_i \lambda^i \;  \;  \; (\lambda \rightarrow 0)
\end{equation}  we get
\begin{equation}
    \begin{gathered}
    \varepsilon_0 = b_0,\,
    \varepsilon_1 = b_0 b_1,\,\varepsilon_2 = b_0{\left(b_1 b_2 +\frac{{b_1 }^2 }{2}\right)},
    \varepsilon_3 = b_0 b_1 {\left( b_2 \,b_3 +\frac{{b_2 }^2 }{2} +b_1b_2+\frac{{b_1 }^2 }{6} \right)},\, \cdots.
\end{gathered}
\end{equation}
Solving these equations sequentially based on the available numerical values of $\varepsilon_i$ of the perturbation series, we get the continued exponential coefficients $b_i$. 

The partial sums for the continued exponential of ground-state energy from Eq.(2) are given by the sequence of approximants \begin{equation}
     A_1 = Z^2a_0\exp(a_1Z^{-1}), A_2 = Z^2a_0\exp(a_1Z^{-1}\exp(a_2Z^{-1})), \cdots. \end{equation} Further, Shanks transformation is used to converge the sequence of partial sums more rapidly. Shanks transformations \cite{bender1999advanced} and their iterations are defined as $$ S(A_i) = \frac{A_{i+1}A_{i-1}-A_i^2}{A_{i+1}+A_{i-1}-2A_i}$$ and $ S^2(A_i) = S(S(A_i)).$ The partial sums for the perturbation series, continued exponential, and their iterated Shanks transformations $S(A_i)$ and $S^2(A_i)$ can be observed in table \ref{table 12} for Z=2. 
    \begin{table}[htp!]
\scriptsize
\begin{center}
\caption{Nonrelativistic ground-state energy of the Helium atom } 

 \begin{tabular}{||c c c c c||}
 
 \hline
 Order ($i$) & Partial sums of perturbation series & $A_i$ & $S(A_i)$ & $S^2(A_i)$ \\ [0.5ex] 
 \hline\hline

 6 & -2.903706946
   & -2.903709215
   & -2.903723819
   & -2.903724371 \\ 
 \hline
 7 & -2.903718576 
   & -2.903721101
   & -2.903724445
   & -2.903724393\\
 \hline
 8 & -2.903722371
   & -2.903723710
   & -2.903724389
   & - \\ 
 \hline
 9 & -2.903723666
   & -2.903724249
   & - 
   & - \\ 
 \hline
\end{tabular}
\label{table 12}
\end{center}
\end{table}

\begin{figure}[htp]
    \centering
    \begin{minipage}[b]{0.52\textwidth}
        \centering
        \includegraphics[width=\textwidth]{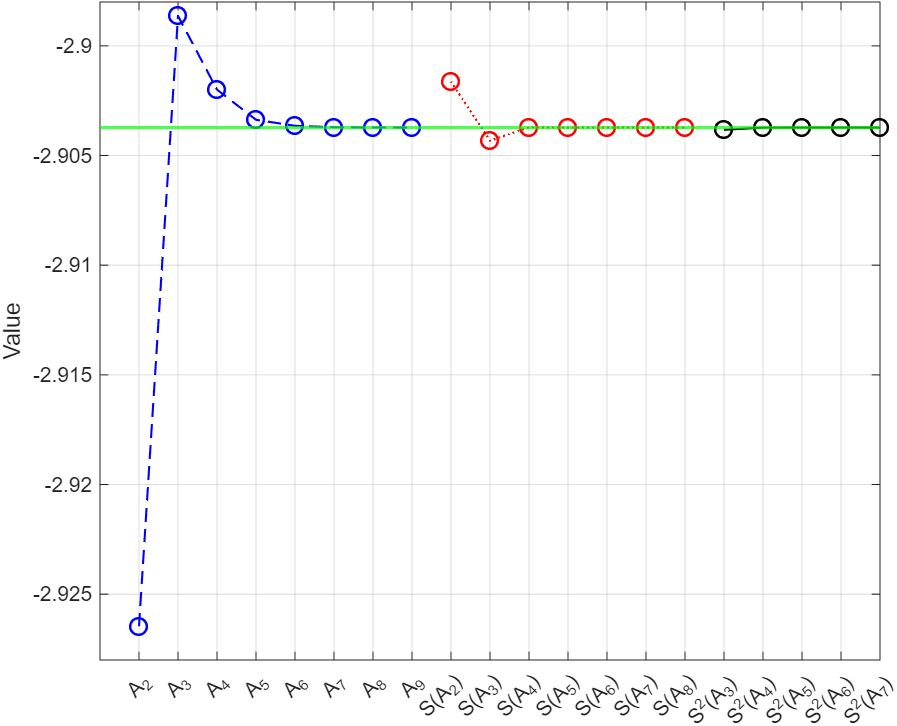} 
    \end{minipage}
    \begin{minipage}[b]{0.54\textwidth}
        \centering
        \includegraphics[width=\textwidth]{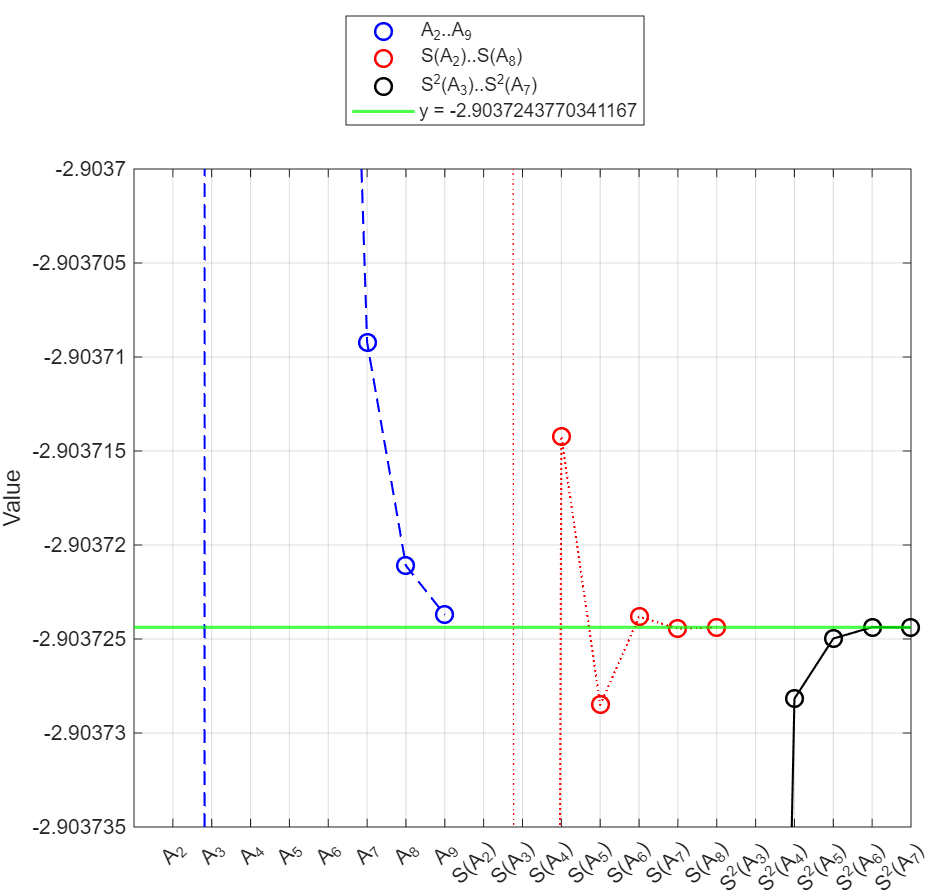} 
    \end{minipage}
    \caption{Estimates of ground-state energy of He atom from continued exponential $A_i$ and its Shanks transforms $S(A_i), S^2(A_i)$ at successive orders compared with result obtained by Hylleraas-Knight-Scherr variational perturbation method \cite{PhysRevA.41.1247} (horizontal green line). The lower image at the smaller scale shows that $A_i$ has convergence, while its Shanks transforms $S(A_i), S^2(A_i)$ show rapid convergence towards the exact value.}
\end{figure}

We can observe that the partial sums of the continued exponential $A_i$ converge better than the partial sums of the perturbation series to the obtained value of ground-state energy of the He atom. And the Shanks $S(A_i)$, along with iterated Shanks $S^2(A_i)$, show rapid convergence towards the exact value. The iterated Shanks transforms $S^2(A_6)$ and $S^2(A_7)$ provide an upper and lower bound to the exact value of the energy previously obtained (PO) with a numerical accuracy of $10^{-9}$. Numerical accuracy measured by calculating $$\left|\frac{\hbox{nearest bound value-PO value}}{\hbox{PO value}} \right|.$$ Similarly, we plot these $A_i$ (Eq. (6)), $ S(A_i)$ and $S^2(A_i)$ at successive orders $i=2,\cdots,9$ in Fig. 1 to explicitly show the convergence behaviour of the transformed continued exponentials in this case.

In a similar manner, we obtain the upper and lower bounds for other He isoelectronic species and compare with standard results obtained from other variational methods \cite{thakkar1994,PhysRevA.65.054501} in table \ref{table 11}. We obtain a numerical accuracy of $10^{-6}$ for $\hbox{H}^-$, $10^{-7}$ for He, $10^{-12}$ for $\hbox{Li}^+$, $\hbox{Be}^{++}$ and $\hbox{B}^{+++}$ ions. 
 \begin{table}[htp!]
 \scriptsize
\begin{center}
\caption{Nonrelativistic ground-state energy of He isoelectronic species} 

 \begin{tabular}{||c c c c||}
 
 \hline
Z & Upper bound & Lower bound & PO value \\ [0.5ex] 
 \hline\hline
 
     1
   & -0.527689
   & -0.527794
   & -0.527751016544377196613 \cite{PhysRevA.65.054501} \\ 
 \hline
     2 
   & -2.903724371
   & -2.903724393 
   & -2.903724377034119598311 \cite{PhysRevA.65.054501} \\
 \hline
     3
   & -7.27991341214
   & -7.27991341307
   & -7.2799134126693020 \cite{thakkar1994} \\ 
 \hline
     4
   & -13.655566238376
   & -13.655566238483
   & -13.6555662384235829 \cite{thakkar1994}\\ 
 \hline
 5
   & -22.030971580257
   & -22.030971580277
   & -22.0309715802427777 \cite{thakkar1994}\\ 
 \hline
\end{tabular}
\label{table 11}
\end{center}
\end{table}
\section{Stark effect in H atom}
The perturbed energy levels of H atom with an electric field in $z$ direction ($\lambda z$) are solved using Hypervirial theorems (HVT) and Hellman-Feynman theorem (HFT) \cite{AUSTIN,fernandezf} for parabolic coordinates. Also, the energy levels are solved for spherical coordinates using the moment method as explained by Fernandez \cite{fernandezf}. The energy eigenvalues are obtained in a perturbation series of $\lambda$ (perturbation parameter expressed in a.u. of electric field, where 1 $ a.u.\approx 5.142\times10^9 \ V/cm$) for any set of quantum numbers ($n_1,n_2,m$). Where the principal quantum number $n$ and magnetic quantum number $m$ are related to the non-negative parabolic quantum numbers $n_1$ and $n_2$, under the relation $n = n_1 + n_2 + |m| +1$. For the $n=1$ ground state, ($0,0,0$), since the odd order coefficients in the perturbation series of $\lambda$ are zero, we convert it into a continued exponential as  
\begin{equation}
    E(\lambda,n_1,n_2,m)=\sum_{i=0}^{\infty} \varepsilon_i\lambda^i \sim b_0\exp(b_1\lambda^2 \exp(b_2 \lambda^2 \exp(b_3 \lambda^2 \exp(b_4 \lambda^2 \exp(\cdots)))))   \label{PS2CE2}
\end{equation} 
 and obtain the approximants \begin{equation}
 B_1 = b_0\exp(b_1\lambda^2),B_2 = b_0\exp(b_1\lambda^2\exp(b_2\lambda^2)),B_3 = b_0\exp(b_1\lambda^2\exp(b_2\lambda^2\exp(b_3\lambda^2))),\cdots. 
\end{equation}  We use $\varepsilon_i$ to obtain $b_i$ up to order $i=16$ and compare standard results in literature with the significant Shanks iteration value, which does not have a pole (table \ref{table 1}). For higher states $n>2$ similarly we convert the perturbation series of $\lambda$ into a continued exponential as  \begin{equation}
    E(\lambda,n_1,n_2,m)=\sum_{i=0}^{\infty} \varepsilon_i\lambda^i \sim c_0\exp(c_1\lambda \exp(c_2 \lambda \exp(c_3 \lambda \exp(c_4 \lambda \exp(\cdots)))))  \label{PS2CE}
\end{equation} 
 and correspondingly use the approximants \begin{equation}
 C_1 = c_0\exp(c_1\lambda),C_2 = c_0\exp(c_1\lambda\exp(c_2\lambda)),C_3 = c_0\exp(c_1\lambda\exp(c_2\lambda\exp(c_3\lambda))),\cdots, \label{approximant}
\end{equation}
Shanks iteration $S^2(C_i)$ to obtain the energy eigenvalues. We obtain $c_i$ up to $i=6$ to compute energy levels of $n=2$, $n=5$, and $n=10$ states and compare with results in the literature in tables \ref{table 2}, \ref{table 3}, and \ref{table 4}, respectively. For $n=1,2,5,10$ energy levels we obtain accuracy of up to $10^{-10}$, $10^{-8}$, $10^{-6}$, $10^{-4}$  for electric fields $\leq$
$10^8 \ V/cm$, $10^6 \ V/cm$, $10^5 \ V/cm$, $10^4 \ V/cm$, respectively. 
\begingroup
\setlength{\tabcolsep}{6pt} 
\renewcommand{\arraystretch}{1} 
\begin{table}[htp!]
\scriptsize
\begin{center}
\caption{Perturbed ground state $n=1$ energy level of H atom in the electric field for varying $\lambda$. We compare our results with literature \cite{FERNANDEZ2018,groundabcd,grounda,variedground,resumm,alexander}.} 

 \begin{tabular}{||c c c||}
 
 \hline
$\lambda$ & \begin{tabular}{c c} Our values \\  $(0,0,0)$  \end{tabular}& \begin{tabular}{c c}  Literature \\  $(0,0,0)$ \end{tabular} \\ [0.5ex] 
 \hline\hline
 0.001
   &  -0.5000022500555518 
   &  -0.5000022500555518 \cite{FERNANDEZ2018} \\
   \hline
     0.005
   & -0.5000562847937930 
   & -0.5000562847937930 \cite{FERNANDEZ2018}
   \\ 
 \hline
     0.025
   & -0.501429291888
   & \begin{tabular}{c c c}   -0.50142929181840 \cite{groundabcd}  \\ -0.50142929180 \cite{grounda} \end{tabular}
  \\
 \hline
     0.03
   & -0.50207427241
   & \begin{tabular}{c c}   -0.50207427260806 \cite{groundabcd} \\ -0.5020742726071 \cite{variedground} \end{tabular}
    \\ 
 \hline
     0.04
   & -0.5037724
   & \begin{tabular}{c c}  
    -0.5037715910136542 \cite{resumm} \\ -0.5037715910137 \cite{variedground} 
  \end{tabular} \\
   \hline
     0.05
   &  -0.50610508
   &  \begin{tabular}{c c}  —0.5061054253626 \cite{variedground} \\ -0.5061054230 \cite{groundabcd} \end{tabular}
   \\ 
   \hline
     0.06
   &  -0.50916
   & \begin{tabular}{c c}  -0.509203450879 \cite{variedground} \\ -0.509203451088 \cite{resumm} \end{tabular}  \\
   \hline
     0.08
   &  -0.5157
   & \begin{tabular}{c c}  
    -0.5175606171  \cite{variedground} \\
     -0.51756050 \cite{resumm} \end{tabular}\\
   \hline
     0.1
   &  -0.5289
   & \begin{tabular}{c c}  -0.527418176 \cite{variedground} \\
   -0.5274193 \cite{resumm}  \end{tabular}\\
   \hline
     0.11
   & -0.537 
   & \begin{tabular}{c c}  -0.535 \cite{grounda}  \\ -0.531 \cite{alexander} \end{tabular}
   \\ 
   \hline
\end{tabular}
\label{table 1}
\end{center}
\end{table}
 \begingroup
\setlength{\tabcolsep}{6pt} 
\renewcommand{\arraystretch}{1} 
\begin{table}[htp!]
\scriptsize
\begin{center}
\caption{Perturbed energy levels of $n=2$ states of H atom in electric field. For each state $(n_1,n_2,m)$ and corresponding $\lambda$, the first line shows the value we calculated, and the second line shows the value from literature \cite{FERNANDEZ2018,Damburg1976}.}

 \begin{tabular}{||c c c c||}
 
 \hline
$\lambda$ & $(0,1,0)$  & $(1,0,0)$ & $(0,0,1)$ \\  [0.5ex] 
 \hline\hline
 
     0.001
  
        &  \begin{tabular}{c c}
             &   -0.128085829\\
             &   -0.1280858350607099 \cite{FERNANDEZ2018}
        \end{tabular} 
        &  \begin{tabular}{c c}
             &  -0.1220826876\\
             &  -0.1220826861326878 \cite{FERNANDEZ2018}
        \end{tabular} 
        &  \begin{tabular}{c c}
             &  -0.1250782240371031\\
             &  -0.1250782240371032 \cite{FERNANDEZ2018}
        \end{tabular} \\
 \hline
     0.004
  
        &  \begin{tabular}{c c}
             &   -0.13847\\
             &   -0.138548793\cite{Damburg1976}
        \end{tabular} 
        &  \begin{tabular}{c c}
             &  -0.11446\\
             &  -0.114305339\cite{Damburg1976}
        \end{tabular} 
        &  \begin{tabular}{c c}
             &  -0.1263170\\
             &  -0.126316885\cite{Damburg1976}
        \end{tabular} \\
 \hline
     0.005
  
        &  \begin{tabular}{c c}
             &   -0.14235\\
             &   -0.1426186075727077 \cite{FERNANDEZ2018}
        \end{tabular} 
        &  \begin{tabular}{c c}
             &   -0.11273\\
             &  -0.1120619240019938 \cite{FERNANDEZ2018}
        \end{tabular} 
        &  \begin{tabular}{c c}
             &   -0.127137\\
             &  -0.1271466127039709 \cite{FERNANDEZ2018}
        \end{tabular} \\
 \hline
 0.008
  
        &  \begin{tabular}{c c}
             &    -0.1555\\
             &   - 0.1563768\cite{Damburg1976}
        \end{tabular} 
        &  \begin{tabular}{c c}
             &   -0.1057\\
             &  -0.1066684\cite{Damburg1976}
        \end{tabular} 
        &  \begin{tabular}{c c}
             &   -0.13149\\
             &  -0.13118859\cite{Damburg1976}
        \end{tabular} \\
 \hline
  0.012
  
        &  \begin{tabular}{c c}
             &     -0.1772\\
             &  -0.171517\cite{Damburg1976}
        \end{tabular} 
        &  \begin{tabular}{c c}
             &   -0.1119\\
             &  -0.100621\cite{Damburg1976}
        \end{tabular} 
        &  \begin{tabular}{c c}
             &  -0.13540\\
             &  -0.1359716\cite{Damburg1976}
        \end{tabular} \\
 \hline
 0.016
  
        &  \begin{tabular}{c c}
             &  -0.199\\
             &  -0.17994\cite{Damburg1976}
        \end{tabular} 
        &  \begin{tabular}{c c}
             &  -0.094\\
             &  -0.092728\cite{Damburg1976}
        \end{tabular} 
        &  \begin{tabular}{c c}
             &   -0.147\\
             &  -0.136437\cite{Damburg1976}
        \end{tabular} \\
 \hline
\end{tabular}
\label{table 2}
\end{center}
\end{table}

\begingroup
\setlength{\tabcolsep}{4pt} 
\renewcommand{\arraystretch}{1} 
\begin{table}[htp!]
\scriptsize
\begin{center}
\caption{Perturbed energy levels of $n=5$ states of the H atom in an electric field. For each state $(n_1,n_2,m)$ and corresponding $\lambda$, the first line shows the value we calculated, and the second and third line shows the value from literature \cite{FERNANDEZ2018,Damburg1976,resumm}.} 

 \begin{tabular}{||c c c | c c||}
 
 \hline
$\lambda \times 10^{4}$ & $(0,4,0)$  &  $(4,0,0)$  & $\lambda\times 10^{4}$ & $(3,0,1)$  \\ [0.5ex] 
 \hline\hline
 
     1
   & \begin{tabular}{c c c}
        &   -0.02317923 \\
        & -0.02317919625030518  \cite{FERNANDEZ2018} \\
        &  -0.0231791962 \cite{Damburg1976}
   \end{tabular}
   & \begin{tabular}{c c c}
        &   -0.017141174 \\
        &   -- \\
        &  --
   \end{tabular}
    & 1.5560
   & \begin{tabular}{c c c}
        & -0.0168572 \\
        & -0.0168552371407617  \cite{resumm} \\
        & -0.0168552372 \cite{Damburg1976}
   \end{tabular}
  \\
 \hline
    1.5
   & \begin{tabular}{c c c}
        &   -0.0249517 \\
        & -0.02495675091807878  \cite{FERNANDEZ2018} \\
        & -0.024956749 \cite{Damburg1976}
   \end{tabular}
   & \begin{tabular}{c c c}
        &  -0.015807706 \\
        & -0.01580776440749585  \cite{FERNANDEZ2018} \\
        &  -0.0158077645 \cite{Damburg1976}
   \end{tabular}
    & 1.9448
   & \begin{tabular}{c c c}
        & -0.0161705 \\
        & -0.0161793882570  \cite{resumm} \\
        & -0.0161793885 \cite{Damburg1976}
   \end{tabular}
  \\
 \hline
    2
   & \begin{tabular}{c c c}
        &    -0.026928 \\
        & -0.02698008147109154  \cite{FERNANDEZ2018} \\
        &  -0.02697136 \cite{Damburg1976}
   \end{tabular}
   & \begin{tabular}{c c c}
        &    -0.0145342 \\
        & -0.01453520517676726  \cite{FERNANDEZ2018} \\
        &  -0.0145352049 \cite{Damburg1976}
   \end{tabular}
    & 2.1393
   & \begin{tabular}{c c c}
        & -0.015840 \\
        & -0.01586046820  \cite{resumm} \\
        & -0.015860468 \cite{Damburg1976}
   \end{tabular}
  \\
 \hline
    2.5
   & \begin{tabular}{c c c}
        &    -0.02921 \\
        & -0.02912946983310681  \cite{FERNANDEZ2018} \\
        &  -0.02896828 \cite{Damburg1976}
   \end{tabular}
   & \begin{tabular}{c c c}
        &    -0.0133211 \\
        & -0.01332892813256598  \cite{FERNANDEZ2018} \\
        &  -0.013328925 \cite{Damburg1976}
   \end{tabular}
    & 2.5282
   & \begin{tabular}{c c c}
        & -0.015205\\
        & -0.015269293  \cite{resumm} \\
        & -0.015269204  \cite{Damburg1976}
   \end{tabular}
  \\
 \hline
    3
   & \begin{tabular}{c c c}
        &    -0.031279 \\
        & -0.03122955458572655  \cite{FERNANDEZ2018} \\
        &  -0.0305381 \cite{Damburg1976}
   \end{tabular}
   & \begin{tabular}{c c c}
        &  -0.012171 \\
        & -0.01220135935615766  \cite{FERNANDEZ2018} \\
        &  -0.01220093 \cite{Damburg1976}
   \end{tabular}
    & 2.9172
   & \begin{tabular}{c c c}
        & -0.01460 \\
        & -0.01474260  \cite{resumm} \\
        &  -0.014740243 \cite{Damburg1976}
   \end{tabular}
  \\
 \hline
    3.5
   & \begin{tabular}{c c c}
        &    -0.03380 \\
        & -0.03323652729915596  \cite{FERNANDEZ2018} \\
        &  -0.0314338 \cite{Damburg1976}
   \end{tabular}
   & \begin{tabular}{c c c}
        &    -0.011086 \\
        & -0.01114288854595917  \cite{FERNANDEZ2018} \\
        &  -0.01113604 \cite{Damburg1976}
   \end{tabular}
    & 3.3061
   & \begin{tabular}{c c c}
        &  -0.01403 \\
        & -0.0142602  \cite{resumm} \\
        &  -0.01424249 \cite{Damburg1976}
   \end{tabular}
  \\
 \hline
    4
   & \begin{tabular}{c c c}
        &    -0.0364 \\
        & -0.03512209724011620  \cite{FERNANDEZ2018} \\
        &  -0.031408 \cite{Damburg1976}
   \end{tabular}
   & \begin{tabular}{c c c}
        &    -0.010070 \\
        & -0.01011729953739499  \cite{FERNANDEZ2018} \\
        &  -0.01008206 \cite{Damburg1976}
   \end{tabular}
    & &
  \\
 \hline
    4.5
   & \begin{tabular}{c c c}
        &   -0.0393 \\
        & -0.03687445248862566  \cite{FERNANDEZ2018} \\
        &  -0.02998 \cite{Damburg1976}
   \end{tabular}
   & \begin{tabular}{c c c}
        &    -0.009025 \\
        & -0.00909725070184054  \cite{FERNANDEZ2018} \\
        &  -0.00899479 \cite{Damburg1976}
   \end{tabular}
    & &
  \\
 \hline
    5
   & \begin{tabular}{c c c}
        &    -0.0423 \\
        & --  \\
        &  -- 
   \end{tabular}
   & \begin{tabular}{c c c}
        &    -0.008096 \\
        & -0.00807076238659657  \cite{FERNANDEZ2018} \\
        &  -0.0078517 \cite{Damburg1976}
   \end{tabular}
    & &
  \\
 \hline
    \end{tabular}
\label{table 3}
\end{center}
\end{table}

\begingroup
\setlength{\tabcolsep}{4pt} 
\renewcommand{\arraystretch}{1} 
\begin{table}[htp!]
\scriptsize
\begin{center}
\caption{Perturbed energy levels of $n=10$ states of H atom in electric field. For each state $(n_1,n_2,m)$ and corresponding $\lambda$, the first line shows the value we calculated, and the second and third line shows the value from literature \cite{FERNANDEZ2018,Kolosov_1987}.} 

 \begin{tabular}{||c c c|c c||}
 
 \hline
$\lambda\times10^5$ & $(9,0,0)$  & $(0,0,9)$
& $\lambda\times10^5$ & $(0,9,0)$ \\  [0.5ex] 
 \hline\hline
 
     2
  
        &  \begin{tabular}{c c c}
             & -2.58593\\
             & -2.585573979364734 \cite{FERNANDEZ2018}\\
             & -2.58557398 \cite{Kolosov_1987}
        \end{tabular} 
        &  \begin{tabular}{c c c}
             & -5.3894\\
             & -5.324404794258087 \cite{FERNANDEZ2018}\\
             & -5.32440479 \cite{Kolosov_1987}
        \end{tabular} 
         & 1.8
        & \begin{tabular}{c c c}
             & -7.7861\\
             & -7.977367228278029  \cite{FERNANDEZ2018}\\
             & -7.977367  \cite{Kolosov_1987}
        \end{tabular} \\
 \hline
      3
  
        &  \begin{tabular}{c c c}
             & -1.5761\\
             & -1.571059822031523 \cite{FERNANDEZ2018}\\
             & -1.57105982 \cite{Kolosov_1987}
        \end{tabular} 
        &  \begin{tabular}{c c c}
             & -5.8301\\
             & -5.648350339949772\cite{FERNANDEZ2018}\\
             & -5.6483507 \cite{Kolosov_1987}
        \end{tabular} 
         & 2.2
        & \begin{tabular}{c c c}
             & -8.6634\\
             & -8.660579416493959 \cite{FERNANDEZ2018}\\
             & -8.660578 \cite{Kolosov_1987}
        \end{tabular} \\
 \hline

\end{tabular}
\label{table 4}
\end{center}
\end{table} 

\subsection{H atom with radial perturbation}
Killingbeck studied the perturbed energy levels in a H atom with radial perturbation $\lambda'r$, using HVT and HFT \cite{KILLINGBECK1978}. He obtained energy levels in the perturbation series of $\lambda'$. Similarly, we convert it into a continued exponential as in eq.(\ref{PS2CE}) and obtain its coefficients from the divergent series of the perturbed ground state energy of the H atom up to $i=9$. The Shanks iterations $S^3(B_5)$ and $S^3(B_6)$ are compared with results obtained using Pad\'e approximants \cite{AUSTIN} in table \ref{table 5}.  
\begingroup
\setlength{\tabcolsep}{6pt} 
\renewcommand{\arraystretch}{1} 
\begin{table}[htp!]
\scriptsize
\begin{center}
\caption{ground-state energy of H in radial perturbation} 

 \begin{tabular}{||c c c c||}
 
 \hline
$\lambda'$ & $S^3(B_5)$ & $S^3(B_6)$ & Pad\'e \cite{AUSTIN} \\ [0.5ex] 
 \hline\hline
 
     0.1
   & -0.36090706
   & -0.36090742
   & -0.36090  \\ 
 \hline
     0.2 
   & -0.23491
   & -0.23567
   & -0.23565 \\
 \hline
     0.3
   & -0.09724
   & -0.11337
   & -0.11892 \\ 
 \hline
     -0.02
   & -0.53066398256
   & -0.53066398238
   & -0.530664  \\ 
 \hline
     -0.03
   & -0.54659128
   & -0.54659132
   & -0.54659 \\ 
    \hline
     -0.04
   & -0.563057
   & -0.563066
   & -0.5631 \\ 
 \hline
\end{tabular}
\label{table 5}
\end{center}
\end{table}
\section{Zeeman effect in H atom}
H atom in a uniform magnetic field is characterized by a coupling parameter $\gamma$ where in a.u. $\gamma=2.3505\times10^9$G. The most accurate perturbation coefficients for the H atom are solved using the moment method as described by Vainberg et al. \cite{Vainberg1998}. We convert the perturbation series of $\gamma$ to a continued exponential and obtain perturbed energy levels for $m=0$, $n=1,2$ states in table \ref{table 9}. For $m=0$, $n=2,3$ the eigenvalues in Rydberg units (=1/2 a.u.) are obtained for $\gamma\leq0.05$ in table \ref{table 10}. For $1s, 2s$ and $2p$, we get high accuracy compared to the existing literature for magnetic fields $\leq 10^8$G, and similarly, for states $3s, 3p$ and $3d$, we get high accuracy for magnetic fields $\leq 10^7$G. 
\begingroup
\setlength{\tabcolsep}{4pt} 
\renewcommand{\arraystretch}{1} 
\begin{table}[htp!]
\scriptsize
\begin{center}
\caption{Perturbed energy levels of $n=1,2$ states of H atom in the magnetic field. For each state and corresponding $\gamma$, the first line shows the value we calculated, and the following lines show the values from the literature.} 

 \begin{tabular}{||c|c|c c||}
 
 \hline
$\gamma$ & $1s$
& $2s$ & $2p$ \\  [0.5ex] 
 \hline\hline
 
     0.0425
  
        &  \begin{tabular}{c c c}
             & -0.4995493\\
             & -0.499548 \cite{peek}\\
             & -0.499548 \cite{Gallas}
        \end{tabular} 
        &  \begin{tabular}{c c c c}
             & -0.11910\\
             & -0.11908 \cite{Cohen_1981}\\
             & -0.1207 \cite{peek}\\
             & -0.118943 \cite{Gallas}
        \end{tabular} 
         & \begin{tabular}{c c c}
             & -0.1124083\\
             & -0.11240 \cite{Cohen_1981}\\
             & -0.11237 \cite{peek}\\
             & -0.112339 \cite{Gallas}
        \end{tabular}
         \\
 \hline
 0.1
  
        &  \begin{tabular}{c c c}
             & -0.4975265\\
             & -0.497525 \cite{Praddaude}\\
             & -0.497512 \cite{Gallas}
        \end{tabular} 
        &  \begin{tabular}{c c c}
             & -0.09829\\
             & -0.098085 \cite{Praddaude}\\
             & -0.095822 \cite{Gallas}
        \end{tabular} 
         & \begin{tabular}{c c c}
             & -0.112427\\
             & -0.112410 \cite{Praddaude}\\
             & -0.111363 \cite{Gallas}
        \end{tabular}
         \\
 \hline
       0.2125
  
        &  \begin{tabular}{c c c}
             & -0.48919\\
             & -0.48906 \cite{peek}\\
             & -0.48892 \cite{Gallas}
        \end{tabular} 
        &  \begin{tabular}{c c c c}
             & -0.03536\\
             & -0.03349 \cite{Cohen_1981}\\
             & -0.05508 \cite{peek}\\
             & -0.02441 \cite{Gallas}
        \end{tabular} 
         & \begin{tabular}{c c c}
             & -0.07306\\
             & -0.07996 \cite{Cohen_1981}\\
             & -0.07758 \cite{peek}\\
             & -0.07378 \cite{Gallas}
        \end{tabular}
         \\
 \hline
  0.425
  
        &  \begin{tabular}{c c c}
             & -0.45998\\
             & -0.45968 \cite{peek}\\
             & -0.45802 \cite{Gallas}
        \end{tabular} 
        &  
         & 
         \\
 \hline
    
\end{tabular}
\label{table 9}
\end{center}
\end{table} 
\begingroup
\setlength{\tabcolsep}{4pt} 
\renewcommand{\arraystretch}{1} 
\begin{table}[htp!]
\scriptsize
\begin{center}
\caption{Perturbed energy levels of $n=2,3$ states of H atom in magnetic field (Rydberg units). For each state and corresponding $\gamma$ the first line shows the value we calculated, following lines show the values from literature.} 

 \begin{tabular}{||c|c c|c c c||}
 
 \hline
$\gamma$ & $2s$
& $2p$ & $3s$ & $3p$ & $3d$ \\  [0.5ex] 
 \hline\hline
 
     0.005
  
        &  \begin{tabular}{c c c}
             & -0.2498251985\\
             & -0.2498 \cite{smith}\\
             & -0.2498 \cite{Gallas}
        \end{tabular} 
        &  \begin{tabular}{c c c c}
             & -0.2499250523\\
             & -0.2499 \cite{smith}\\
             & -0.2499 \cite{Gallas}
        \end{tabular} 
         &  \begin{tabular}{c c c c}
             & -0.1101418831\\
             & -0.1101 \cite{smith}\\
             & -0.1102 \cite{Gallas}
        \end{tabular} 
        &  \begin{tabular}{c c c c}
             & -0.1106654213\\
             & -0.1106 \cite{smith}\\
             & -0.1106 \cite{Gallas}
        \end{tabular} 
        &  \begin{tabular}{c c c c}
             & -0.1108537891\\
             & -0.1108 \cite{smith}\\
             & -0.1107 \cite{Gallas}
        \end{tabular} 
         \\
 \hline
 0.02
  
        &  \begin{tabular}{c c c}
             & -0.2472483551\\
             & -0.2472 \cite{smith}\\
             & -0.2472 \cite{Gallas}
        \end{tabular} 
        &  \begin{tabular}{c c c c}
             & -0.2488129320\\
             & -0.2488 \cite{smith}\\
             & -0.2488 \cite{Gallas}
        \end{tabular} 
         &  \begin{tabular}{c c c c}
             & -0.097279\\
             & -0.09727 \cite{smith}\\
             & -0.0986 \cite{Gallas}
        \end{tabular} 
        &  \begin{tabular}{c c c c}
             & -0.104757\\
             & -0.1047 \cite{smith}\\
             & -0.1043 \cite{Gallas}
        \end{tabular} 
        &  \begin{tabular}{c c c c}
             & -0.107244\\
             & -0.1072 \cite{smith}\\
             & -0.1053 \cite{Gallas}
        \end{tabular} 
         \\
 \hline
       0.05
  
        &  \begin{tabular}{c c c}
             & -0.2340335\\
             & -0.2340 \cite{smith}\\
             & -0.2335 \cite{Gallas}
        \end{tabular} 
        &  \begin{tabular}{c c c c}
             & -0.2429297\\
             & -0.2429 \cite{smith}\\
             & -0.2427 \cite{Gallas}
        \end{tabular} 
         &  \begin{tabular}{c c c c}
             & -0.05587\\
             & -0.04939 \cite{smith}\\
             & -0.05025 \cite{Gallas}
        \end{tabular} 
        &  \begin{tabular}{c c c c}
             & -0.08365\\
             & -0.08276 \cite{smith}\\
             & -0.0755 \cite{Gallas}
        \end{tabular} 
        &  \begin{tabular}{c c c c}
             & -0.09195\\
             & -0.09096 \cite{smith}\\
             & -0.08063 \cite{Gallas}
        \end{tabular} 
         \\
 \hline

\end{tabular}
\label{table 10}
\end{center}
\end{table} 
\section{Conclusion}
We have obtained eigenvalues in certain perturbation problems of atomic systems and compared them with literature from different variational calculations. This reassures the perturbational approach taken to address these physical systems. It can generally be observed that accuracy reduces for higher perturbation parameters, which can be increased by taking higher perturbation coefficients within a specific parameter region. We take only ten or fewer perturbation coefficients to keep the method computationally simple. The higher coefficients in the Taylor expansion of the continued exponential are long and require more computational capability. We implement a simple method to obtain meaningful answers from divergent perturbation series for small perturbation parameters and find an approximate region to which
the method is applicable in these instances. Furthermore, continued exponential functions can be tried to obtain convergence behavior in a wide range of perturbation methods. 

While the present work has focused on continued exponentials as a simple tool without free parameters to handle divergent perturbation series in atomic systems, it is helpful to view this approach as one member of a broader class of continued function resummation techniques.  In particular, continued fractions and related Padé approximant methods thus have broad applicability across physics. They systematically improve the convergence of perturbation series and turn formal expansions into useful analytic functions \cite{BAKER1970,Baker1981}. For eigenvalue problems such as for bound-states of the Schrödinger equation, specialized continued-fraction recurrences yield stable, high-precision spectra where straightforward polynomial methods fail \cite{GERCK1980}. In fluid dynamics, Padé-based resummation extends flow-series solutions far beyond their naive convergence limits. Such rational approximations have been applied in both steady and unsteady, incompressible and compressible regimes, effectively capturing flow behavior that straightforward Taylor expansions miss \cite{pozzi1994applications}. These ideas also permeate nonlinear wave analysis, where recent studies of variable-coefficient wave equations \cite{Zhang2024,Lu2024}, exploit Painlevé analysis for checking the analytic properties, which is closely related to Padé-based techniques \cite{Novokshenov2009}. These examples demonstrate that continued function techniques provide a unifying framework for taming divergent series and obtaining analytic approximations in quantum physics, fluid dynamics, and nonlinear systems.

\section*{Declaration}
No funds, grants, or other support were received to the author.






\bibliographystyle{ieeetr}
\bibliography{sample.bib}







\end{document}